\documentclass[twocolumn,showpacs,superscriptaddress,preprintnumbers,prd]{revtex4}
\usepackage{graphicx,amsmath,dcolumn,bm}
\begin{document}
\preprint{KEK-TH-1357}
\title{Tensor-polarized quark and antiquark distribution functions
       in a spin-one hadron}
\author{S. Kumano}
\affiliation{KEK Theory Center,
          Institute of Particle and Nuclear Studies \\
          High Energy Accelerator Research Organization (KEK) \\
            and Department of Particle and Nuclear Studies \\
             Graduate University for Advanced Studies \\
           1-1, Ooho, Tsukuba, Ibaraki, 305-0801, Japan}     
\date{June 14, 2010}
\begin{abstract}
To understand orbital-angular-momentum contributions is becoming
crucial for clarifying nucleon-spin issue in the parton level.
Twist-two structure functions $b_1$ and $b_2$ for spin-one hadrons 
could probe orbital-angular-momentum effects, which reflect 
a different aspect from current studies for the spin-1/2 nucleon,
since they should vanish if internal constituents are in the $S$ state.
These structure functions are related to tensor structure
in spin-one hadrons. Studies of such tensor structure will open
a new field of high-energy spin physics. The structure functions $b_1$
and $b_2$ are described by tensor-polarized quark and antiquark 
distributions $\delta_T q$ and $\delta_T \bar q$. 
Using HERMES data on the $b_1$ structure function for the deuteron,
we made an analysis of extracting the distributions $\delta_T q$ and 
$\delta_T \bar q$ in a simple $x$-dependent functional form. 
Optimum distributions are proposed for the tensor-polarized
valence and antiquark distribution functions from the analysis.
A finite tensor polarization is obtained for antiquarks if we impose
a constraint that the first moments of tensor-polarized valence-quark
distributions vanish. It is interesting to investigate a physics
mechanism to create a finite tensor-polarized antiquark distribution.
\end{abstract}
\maketitle


Origin of nucleon spin has been investigated extensively after 
the EMC discovery that almost none of nucleon spin is carried 
by quarks \cite{emc88}. Recent studies of polarized parton 
distribution functions (PDFs) are found in Ref. \cite{ppdfs}.
Although a gluon-spin contribution is not determined accurately,
orbital angular momenta are likely to be the crucial quantities 
in explaining the nucleon spin. Such contributions have been
investigated recently by generalized parton distribution functions
in lepton scattering \cite{recent-gpds} and will be studied
possibly at hadron facilities \cite{hadron-gpds}. 

There are other quantities which are sensitive to the orbital
angular momenta. For example, there are twist-two structure 
functions $b_1$ and $b_2$ in spin-one hadrons \cite{fs83,hjm89}.
They could be related to the orbital angular momenta of internal
constituents because they vanish if the constituents are
in the $S$ wave. Of course, they probe a different aspect
of orbital-angular-momentum effects from the current ones 
for the nucleon because they are related to tensor-structure
nature of spin-one hadrons. 
It is noteworthy that tensor structure is not understood at all
in the parton level, which suggests that a new field of
spin physics should be created by investigating
the tensor-polarized structure functions.

New polarized structure functions ($b_1$, $b_2$, $b_3$, and $b_4$) 
were introduced in describing lepton deep inelastic scattering 
from a spin-one hadron \cite{fs83,hjm89,higher-spin}.
A useful sum rule for the twist-two function $b_1$ was
proposed in Ref. \cite{b1-sum}, and it is partially used 
in this work.
In conventional hadron models, such tensor structure arises 
due to the $D$-state admixture \cite{fs83,hjm89,kh91}, 
pions \cite{miller-b1}, and shadowing effects 
\cite{b1-shadowing} if the target is the deuteron.
However, the tensor structure would not be simply 
described by such conventional models at high energies according
to our experience on the nucleon-spin issue.
On the other hand, a theoretical formalism was developed in 
Ref. \cite{pd-drell-yan} to investigate the tensor-polarized
distributions at hadron facilities by Drell-Yan processes
with polarized deuteron.
There are related theoretical studies such as new fragmentation 
functions \cite{spin-1-frag}, generalized parton distributions 
\cite{spin-1-gpd}, target mass corrections \cite{mass-corr},
positivity constraints \cite{dmitrasinovic-96},
lattice QCD estimate \cite{lattice},
projection operators of $b_{1-4}$ \cite{kk08} for spin-one hadrons.
The first measurement of the structure function $b_1$ was
done by the HERMES collaboration in 2005 \cite{hermes05}. 

The purpose of this research is to propose a simple parametrization
for the tensor-polarized quark and antiquark distribution functions 
by analyzing the HERMES data. It is intended to understand
the current status of the tensor distributions. Obtained 
distributions could be used for comparing them with theoretical 
model calculations and for proposing future experiments. 

The structure function $b_1$ is defined in the hadron tensor
$W_{\mu\nu}$ \cite{hjm89,kk08}. 
It is expressed in term of tensor-polarized distributions 
($\delta_T q$ and $\delta_T \bar q$)
as \cite{hjm89,b1-sum,factor-2}
\begin{equation}
b_1 (x,Q^2) = \frac{1}{2} \sum_i e_i^2 
   \left[ \delta_T q_i(x,Q^2) 
        + \delta_T \bar q_i(x,Q^2) \right ] ,
\end{equation}
where $i$ indicates the flavor of a quark and $e_i$ is the charge
of the quark. The variables $Q^2$ and $x$ are defined 
by the momentum transfer $q$ as $Q^2=-q^2$
and $x=Q^2/(2 M_N \nu)$, where $M_N$ and $\nu$ are the nucleon 
mass and the energy transfer, respectively. 
Hereafter, the $Q^2$ dependence is not explicitly
written in the PDFs. In this work, the $b_1$, $\delta_T q_i$,
$\delta_T \bar q_i$, and unpolarized PDFs are defined by the ones 
{\it per nucleon} for a nuclear target, namely they are 
divided by the factor of two if it is the deuteron.
The tensor-polarized distribution $\delta_T q$ is defined by
\begin{equation}
\delta_T q_i (x) \equiv q^0_i (x)
    -\frac{q^{+1}_i (x) +q^{-1}_i (x)}{2}  , 
\end{equation}
where $q_i^\lambda$ indicates an unpolarized-quark distribution
in the hadron spin state $\lambda$, and it is also defined
the one per nucleon. Namely, $\delta_T q$ indicates
an unpolarized-quark distribution in a tensor-polarized
spin-one hadron. It should be noted that the notation $\delta_T q$
is not the transversity distribution, for which similar notations
($\delta$ or $\Delta_T$) are used in nucleon-spin studies,
throughout this article.
A sum rule exists for $b_1$ in a parton model \cite{b1-sum}:
\begin{equation}
\int dx \, b_1 (x)  = - \frac{5}{24} 
           \lim_{t \rightarrow 0} t F_Q (t) = 0 ,
\label{eqn:b1-sum}
\end{equation}
if the tensor-polarized antiquark distributions vanish
$\int dx \, \delta_T \bar q (x) = 0$.
Here, $F_Q(t)$ is the electric quadrupole form factor of 
a spin-one hadron at the momentum squared $t$.

We analyze the HERMES $b_1$ data for the deuteron.
The tensor-polarized distributions are introduced as the unpolarized 
PDFs in the deuteron ($D$) multiplied by a common weight function
$\delta_T w(x)$:
\begin{align}
\delta_T q_{iv}^D (x) & = \delta_T w(x) \, q_{iv}^D (x), 
\nonumber \\
\delta_T \bar q_i^D (x) & 
           = \alpha_{\bar q} \, \delta_T w(x) \, \bar q_i^D (x) .
\label{eqn:delta-q-qbar-1}
\end{align}
Namely, certain fractions of quark and antiquark distributions
are tensor polarized and such probabilities are given by 
the function $\delta_T w(x)$ and an additional constant $\alpha_{\bar q}$
for antiquarks in comparison with the quark polarization.
The $x$ dependence of $\delta_T w(x)$ for antiquarks could be different
in general from the one for quarks. However, it is not the stage
of suggesting such a difference from experimental measurements
as it will become obvious later in this article.

It is known that nuclear modifications are less than
a few percent for the unpolarized PDFs in the deuteron \cite{hkn07}.
The tensor-polarized distributions cannot be determined
within a few percent accuracy at this stage. Therefore,
the nuclear modifications are neglected in $q_i^D$ 
and $\bar q_i^D$. Then, the PDFs in the deuteron
are written by a simple addition of proton and neutron
contributions: $q_i^D=(q_i^p+q_i^n)/2$ and 
$\bar q_i^D=(\bar q_i^p+\bar q_i^n)/2$.
Furthermore, isospin symmetry is assumed for relating the PDFs
of the neutron to the ones of the proton:
$u_n=d$, $d_n=u$, $\bar u_n=\bar d$, and $\bar d_n=\bar u$.
Then, the tensor polarized distributions are
\begin{align}
\delta_T q_v^D (x) & \equiv \delta_T u_v^D (x) 
                         = \delta_T d_v^D (x) 
\nonumber \\
\    & = \delta_T w(x) \, \frac{u_v (x) +d_v (x)}{2} , 
\nonumber \\
\delta_T \bar q^D (x) & \equiv \delta_T \bar u^D (x)
                            = \delta_T \bar d^D (x)
                            = \delta_T      s^D (x)
                            = \delta_T \bar s^D (x)                    
\nonumber \\
\    & = \alpha_{\bar q} \, \delta_T w(x) \, 
    \frac{2 \bar u(x) +2 \bar d(x) +s(x) + \bar s(x)}{6} ,
\label{eqn:dw(x)}
\end{align}
where flavor-symmetric tensor-polarized antiquark distributions
are assumed. Tensor polarized heavy-quark distributions are 
neglected in this work.
To be precise, the distributions extend to $x=2$ in the deuteron,
whereas the kinematical limit is $x=1$ for the nucleon. 
Therefore, the tensor-polarized distributions given in 
Eq. (\ref{eqn:dw(x)}) cannot describe the region at $1<x<2$. 
However, the PDFs are very small and it is not the stage
to investigate the tensor distributions in such a large-$x$ region.

\begin{table*}[t]
\caption{Determined parameters in Eqs. 
(\ref{eqn:dw(x)}) and (\ref{eqn:dw(x)-abc}).
$Q^2$ is taken $Q^2$=2.5 GeV$^2$.}
\label{parameter-delta-q}
\centering
\begin{tabular}{@{\hspace{0.2cm}}c|@{\hspace{0.4cm}}c@{\hspace{0.4cm}}
c@{\hspace{0.4cm}}c@{\hspace{0.4cm}}c@{\hspace{0.4cm}}
c@{\hspace{0.4cm}}c@{\hspace{0.2cm}}}
\hline
Analysis  &  $\chi^2$/d.o.f.    &  $a$                
          &  $\alpha_{\bar q}$  &  $b$  
          &  $c$                &  $x_0$              \\
\hline
Set 1     &  2.83               &  0.378 $\pm$ 0.212     
          &  0.0 (fixed)        &  0.706 $\pm$ 0.324 
          &  1.0  (fixed)       &  0.229              \\
Set 2     &  1.57               &  0.221 $\pm$ 0.174 
          &  3.20 $\pm$ 2.75    &  0.648 $\pm$ 0.342
          &  1.0  (fixed)       &  0.221              \\
\hline
\end{tabular}
\end{table*}

We analyze the data in the leading order (LO) of the running
coupling constant $\alpha_s$.
The structure function $b_1$ is then given by
\begin{align}
b_1^D (x) & = \frac{1}{36} \delta_T w(x) \, \left [ \,
     5 \{ u_v (x) + d_v (x) \}   \right.
\nonumber \\ 
& \left.
  +4 \alpha_{\bar q} \{ 2 \bar u (x) + 2 \bar d (x)
  +   s (x) + \bar s (x) \} \, \right ] \, .
\label{eqn:b1x}
\end{align}
The unpolarized PDFs $u_v(x)$, $d_v(x)$, $\cdot\cdot\cdot$, $\bar s(x)$
could be taken from a recent global analysis, for example, by 
CTEQ \cite{cteq6.6}, GJR \cite{gjr08}, or MSTW \cite{mstw08}. 
In this work, the LO version of the MSTW parametrization
is used.
For the functional form of $\delta_T w(x)$, we note that
there is a constraint from the sum rule in Eq. (\ref{eqn:b1-sum}).
The integrated tensor polarization, namely the first moment, 
should vanish for the valence quarks.
It indicates that there should be a node in the $x$-dependent function,
so that an appropriate parametrization could be
\begin{equation}
\delta_T w(x) = a x^b (1-x)^c (x_0-x) ,
\label{eqn:dw(x)-abc}
\end{equation}
where $x_0$ is the position where $\delta_T q_v(x)$ 
(and $\delta_T \bar q(x)$) vanishes. 
If the first moments vanish for the valence-quark distributions,
the constant $x_0$ is expressed by the other parameters as
\begin{equation}
x_0= \frac{\int_0^1 dx x^{b+1} (1-x)^c 
          \{ u_v (x) + d_v (x) \} }
          {\int_0^1 dx x^{b} (1-x)^c 
          \{ u_v (x) + d_v (x) \} } .
\end{equation}
The $a$, $b$, $c$, and $\alpha_{\bar q}$ are the parameters to be 
determined from experimental measurements.

The parametrization of Eq. (\ref{eqn:dw(x)-abc}) is motivated
by the following considerations. First, the parton model
indicates the existence of a node as mentioned.
Next, we expect to have smooth polynomial functional forms
in the limits, $x \rightarrow 0$ and $1$, as usual in unpolarized
and longitudinally-polarized PDFs. In addition, the existence of 
the node and the functional form are, for example, supported
by theoretical estimates of a convolution model, or so called 
binding model, where the $D$-state admixture gives rise to 
an $x$-distribution with a node in $b_1$ including 
the tensor-polarized antiquark distributions \cite{fs83,hjm89,kh91}.

From an analysis of the HERMES experimental data, the optimum function
$\delta_T w (x)$ and $\alpha_{\bar q}$ are determined. It is obvious
from the data that the $b_1$ structure functions are not accurately
measured to discuss scaling violation or even details of 
$x$ dependence. Therefore, a simplification is made by ignoring
the scaling violation. The $Q^2$ value is fixed at
$Q^2$=2.5 GeV$^2$, which is about the average $Q^2$ of the HERMES
measurements, for calculating the unpolarized PDFs 
\cite{mstw08} in Eqs. (\ref{eqn:dw(x)}) and (\ref{eqn:b1x}).
We made two types of analyses:
\begin{itemize}
\item Set 1: Tensor-polarized antiquark distributions are terminated
             ($\alpha_{\bar q}$=0).
\item Set 2: Finite tensor-polarized antiquark distributions
             are allowed ($\alpha_{\bar q}$ is a parameter).
\end{itemize}
Due to the lack of data at large $x$, the parameter $c$ cannot
be determined from the current data. We checked that the $\chi^2$ value
is not much affected by this parameter.
Therefore, it is fixed at $c=1$ in our analyses. 

\begin{figure}[b]
\begin{center}
\includegraphics[width=0.40\textwidth]{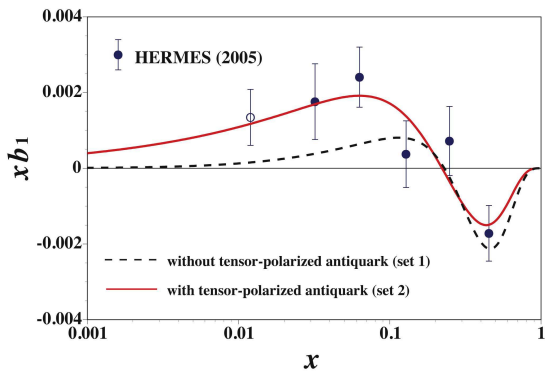}
\vspace{-0.3cm}
\caption{(Color online) Comparison with HERMES data \cite{hermes05}.
The solid and dashed curves indicate theoretical results
with ($\alpha_{\bar q} \ne 0$) and without ($\alpha_{\bar q} = 0$) 
tensor-polarized antiquark distributions.
The open circle is the data at $Q^2<1$ GeV$^2$.}
\label{fig:xb1}
\end{center}
\end{figure}

The determined parameters are listed in Table \ref{parameter-delta-q}.
It is obvious that the fit is not good enough ($\chi^2$/d.o.f.=2.83)
if the tensor-polarized antiquark distributions are terminated 
(set 1) by fixing the parameter as $\alpha_{\bar q}=0$.
If we let this parameter free in the analysis, it is a reasonably
successful one  ($\chi^2$/d.o.f.=1.57).
It is interesting to find that the parameter $\alpha_{\bar q}$ is
larger than one, which indicates that a significant 
tensor polarization exists in the antiquark distributions.

\begin{figure}[b]
\begin{center}
\includegraphics[width=0.40\textwidth]{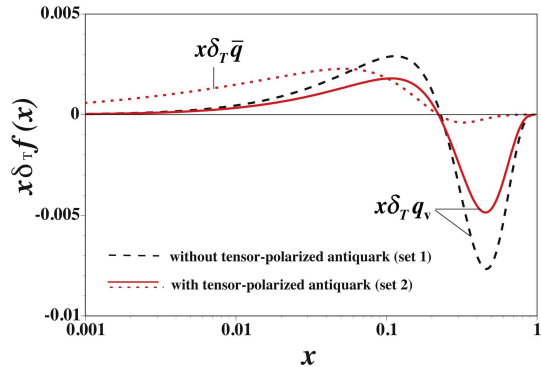}
\vspace{-0.3cm}
\caption{(Color online) Determined tensor-polarized distributions.
The dashed and solid curves are the valence-quark distributions
$x \delta_T q_v^D$ in the deuteron for set 1 ($\alpha_{\bar q} = 0$)
and set 2 ($\alpha_{\bar q} \ne 0$), respectively, 
and the dotted curve is the antiquark distribution 
$x \delta_T \bar q^D$ of set 2.}
\label{fig:x-delta-fx}
\end{center}
\end{figure}

Analysis results are shown in Fig. \ref{fig:xb1} in comparison
with the HERMES experimental data. Only the data with $Q^2>1$ GeV$^2$
are included in the analyses. The set-1 curve is shown by the dashed
curve, which does not agree with the data in the small-$x$ region
($x<0.1$) without the antiquark polarization. The overall fit
is successful only if the antiquark polarization is introduced
(set 2) as shown by the solid curve.
Of course, the results depend on the assumed functional form
including the assumption of using the common weight function 
$\delta_T w(x)$ for the quark and antiquark distributions.
However, it would be reasonable as long as a smooth $x$-dependence
is valid for the weight function $\delta_T w(x)$. 

The determined tensor-polarized distributions in Eq. (\ref{eqn:dw(x)})
are shown in Fig. \ref{fig:x-delta-fx} by using the parameters in
Table \ref{parameter-delta-q}. The amplitude is sightly larger
for the valence-quark distribution of set 1 because the antiquark
distributions are terminated by setting $\alpha_{\bar q}=0$.
The antiquark distribution $\delta_T \bar q^D$ is shown by
the dotted curve and it is mainly distributed in the region $x<0.1$.
It is interesting to find that a finite antiquark tensor
polarization is needed to explain the HERMES data on $b_1$.
If its effect on the $b_1$ sum rule is estimated, we obtain
\begin{align}
\int dx \, b_1 (x) & = - \frac{5}{24} \lim_{t \rightarrow 0} t F_Q (t)
  + \frac{1}{18} \int dx \, [ \, 8 \delta_T \bar u (x) 
\nonumber \\
& \ \ \ \ \ \ 
    + 2 \delta_T \bar d(x)  +\delta_T s (x) + \delta_T \bar s(x) \, ]  
\nonumber \\
& = 0.0058 .
\label{eqn:b1x-sum-value}               
\end{align}
The choice of parametrization of Eq. (\ref{eqn:dw(x)-abc})
for the antiquark distributions could affect the numerical result.
However, as it is obvious from Figs. \ref{fig:xb1}
and \ref {fig:x-delta-fx}, the antiquarks contribute only
at small $x$ ($x<0.1$). As long as the function $\delta_T w(x)$
is a smooth function at $x<0.1$, the result is not 
significantly changed.

This work is the first attempt to parametrize the tensor
polarized valence-quark and antiquark distributions. 
Including the antiquark tensor polarization, we obtained
much smaller $\chi^2$/d.o.f. and it led to a finite
sum as shown in Eq. (\ref{eqn:b1x-sum-value}).
This is a new and interesting result which needs to be
explained theoretically.
The integral is compared with the HERMES results \cite{hermes05},
$\int_{0.002}^{0.85} dx b_1(x) =
   [1.05 \pm 0.34 \text{ (stat)} \pm 0.35 \text{ (sys)}]
   \times 10^{-2}$
and
$\int_{0.02}^{0.85} dx b_1(x) =
   [0.35 \pm 0.10 \text{ (stat)} \pm 0.18 \text{ (sys)}]
   \times 10^{-2}$
in the restricted range with $Q^2>1$ GeV$^2$.
The integral of Eq. (\ref{eqn:b1x-sum-value}) is similar
to the Gottfried sum \cite{flavor3}
\begin{align}
& \int \frac{dx}{x} \, [F_2^p (x) - F_2^n (x) ] = 
   \frac{1}{3}
  +\frac{2}{3} \int [ \bar u(x) - \bar d(x) ] ,
\end{align}
where the deviation from $\int [ u_v(x) - d_v (x) ]/3=1/3$
indicates flavor asymmetric antiquark distributions.
In the $b_1$ case, the finite sum $\int dx b_1$ suggests 
that a finite tensor-polarized antiquark distribution should exist.

It is obvious from Fig. \ref{fig:xb1} that much better measurements
are needed to investigate the details of tensor-polarized 
distributions particularly at medium and large $x$ ($>0.2$). 
Such measurements could be possible, for example, 
at JLab (Thomas Jefferson National Accelerator Facility)
by measuring $b_1$ and also at hadron facilities such as
J-PARC (Japan Proton Accelerator Research Complex) \cite{j-parc}
GSI-FAIR (Gesellschaft f\"ur Schwerionenforschung -Facility for 
Antiproton and Ion Research) \cite{gsi-fair}
by Drell-Yan processes with polarized deuteron \cite{pd-drell-yan}.
In particular, the Drell-Yan processes are suitable for 
directly finding the tensor-polarized
antiquark distributions in Eq. (\ref{eqn:b1x-sum-value}).

{\it Summary}:
In this work, optimum tensor-polarized quark and antiquark distributions
are proposed from the analyses of HERMES data on $b_1$ for the deuteron.
We found that a significant antiquark tensor polarization exists
if the overall tensor polarization vanishes for the valence quarks
although such a result could depend on the assumed functional form.
Further experimental measurements are needed for $b_1$ such 
as at JLab as well as Drell-Yan measurements with tensor-polarized
deuteron at hadron facilities, J-PARC and GSI-FAIR. On the other hand,
it is interesting to conjecture a possible physics mechanism to create
a finite tensor-polarized antiquark distribution.

\vspace{-0.2cm}
\begin{acknowledgements}
\vspace{-0.3cm}
The author thanks 
 C. Ciofi degli Atti, T.-Y. Kimura, Y. Miyachi, and O. V. Teryaev 
for discussions on structure functions of spin-one hadrons.
\end{acknowledgements}

\vspace{-0.0cm}



\begin{thebibliography}{00}
\bibitem{factor-2} The overall factor 1/2 is introduced in $b_1$
            as usual in defining $F_1$ and $g_1$ in terms of PDFs.
\end{thebibliography}
\end{document}